# Technology Transfer and the End of the Bayh-Dole Effect:

# Patents as an Analytical Lens on University-Industry-Government Relations

Loet Leydesdorff [a] & Martin Meyer [b]

**Abstract**

Three periods can be distinguished in university patenting at the U.S. Patent and Trade Office (USPTO) since the Bayh-Dole Act of 1980: (1) a first period of exponential increase in university patenting till 1995 (filing date) or 1999 (issuing date); (2) a period of relative decline since 1999; and (3) in most recent years—since 2008—a linear increase in university patenting. We argue that this last period is driven by specific non-US universities (e.g., Tokyo University and Chinese universities) patenting increasingly in the U.S.A. as the most competitive market for high-tech patents.

[a] University of Amsterdam, Amsterdam School of Communication Research (ASCoR), Kloveniersburgwal 48, 1012 CX Amsterdam, The Netherlands; loet@leydesdorff.net ; http://www.leydesdorff.net
[b] Kent Business School, University of Kent, Canterbury, Kent, CT2 7PE, UK, Katholieke Universiteit Leuven, ECOOM Research Centre for R&D Monitoring, Leuven, Belgium; University of Vaasa, SC-Research, Lapua. Finland.

**Introduction**

Etzkowitz (in press) is publishing, in our opinion prematurely, our Figure 1 entitled "Percentage share of USPTO granted patents to Universities and Institutes of Technology" in a commentary to Figure 2 in Leydesdorff & Meyer (2010: 358), a study in which we reported on the decline of university patenting as signaling possibly "the end of the Bayh-Dole Effect" in the USA. However, we had hitherto deliberately abstained from publishing this new figure (which was presented at conferences) because we thought it prudent to wait until the data for 2012 could be included (cf. Mowery & Sampat, 2004:120; Wong & Singh, 2010). We collected this data in January 2013. Indeed, the new trend can be confirmed: the fit of the upward linear regression since 2008 has now further increased to $R^2 = .98$ (Figure 1). However, the Bayh-Dole effect in the period from 1980-1999 can be considered as a phase of exponential growth ($R^2 > .99$).

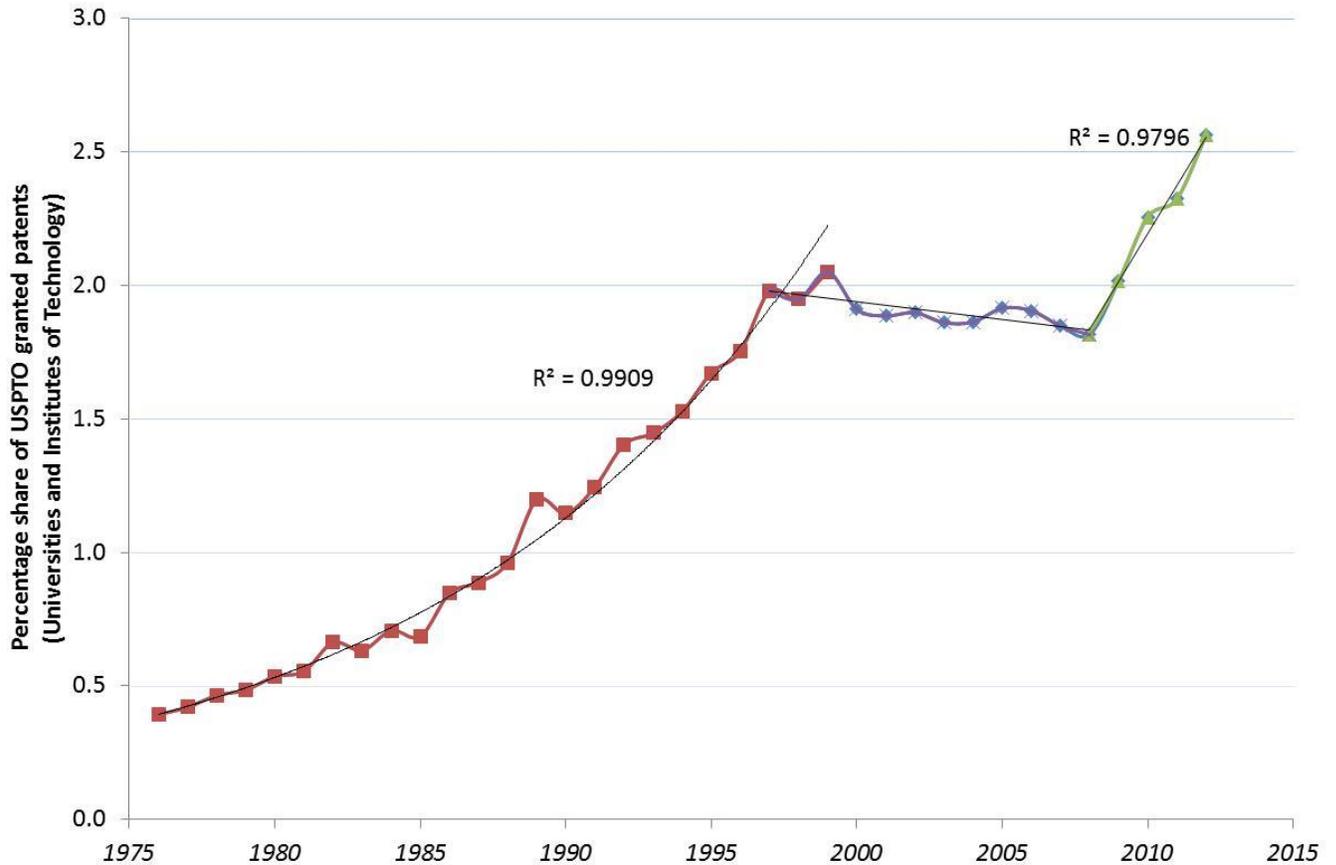

**Figure 1**: Long-term trend of US patenting by Universities and Institutes of Technology; issue dates.

We are currently analyzing this recent trend-breach in greater detail (for a presentation at the Triple Helix Conference in London, July 2013); but under the pressure of Etzkowitz' comments, we provide some preliminary results of our analysis in this rejoinder. It should be notified that these patent statistics are based on issue dates and not on filing dates, as is more common in the management literature (Jaffe & Trajtenberg, 2002). Because delays in the application process are different on average for universities and industries, the current trend happens to be decreasing instead of increasing when measured with filing dates. In other words, filing dates cannot be used for patent statistics data for the most recent years because of the effects of delays at the patent offices.

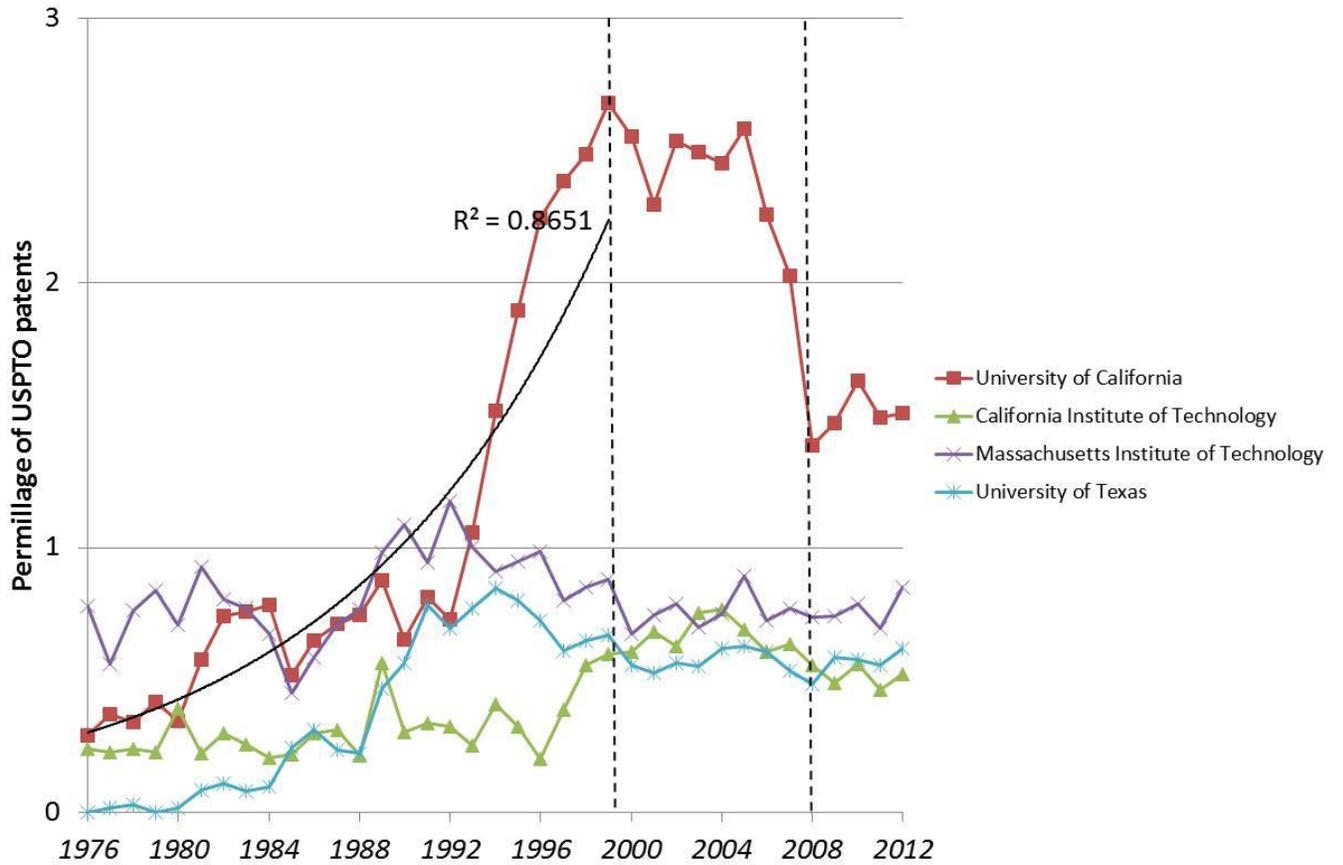

**Figure 2**: Permillage of US patenting for four major American universities and Institutes of Technology.

Figure 2 extends our previous Figure 3 (Leydesdorff & Meyer, 2010: 359) for four major American universities and Institutes of Technology to 2012 using USPTO data. We have added dashed lines at 1999 and 2008 in order to distinguish the following three periods in the data (based on Figure 1):

1. The exponential increase in university patenting during the period from 1980-1995 caused (?) by the Bayh-Dole Act of 1980; using filing dates the curve peaks in 1995, but with issue dates in 1999;
2. A period of relative decline from 1999 until recently;

3. A recent increase since 2008. Since these are (delayed) issue dates, this increase cannot be attributed solely to the financial crisis of 2008; one should look for other possible causes.

The Regents of the University of California administer a number of campuses and laboratories, and thus the general trends are best reflected in this larger data set. The growth period (1976-1999; issue dates) fits roughly to an exponential curve ($R^2 = .87$), and one can see that the sharp decline in the second period has come to a halt since 2008. Perhaps, one can even argue for some increase thereafter, but this tendency seems non-significant. Adding a curve to Figure 1 for only US addresses shows a similar pattern (not included here). Thus, the recent rise in university patenting did not originate in mainly American universities.

**"Inverse foreign direct investment"**

After examining this data in more detail, we feel confident in suggesting that this recent increase is an effect of what one could call "inverse foreign investment" in US patenting. China, for example, is among the assignee countries for more than four percent of the university patents at the USPTO in 2012, as against only 1.8% in 2007. University patenting is an important component of the Chinese portfolio because the division of labour between universities and industry is different in China.

Specific universities, notably foreign ones, began to file for patents at the USPTO as the most competitive market place for new technologies during the second period distinguished above. In the most recent period, Figure 3 shows, for example, a rise for Oxford University that has established a specific firm—ISIS Innovation—for patenting, whereas Cambridge University has retreated from this competition. Cambridge University, however, has remained a major center of developments in both biotechnology and regional developments.

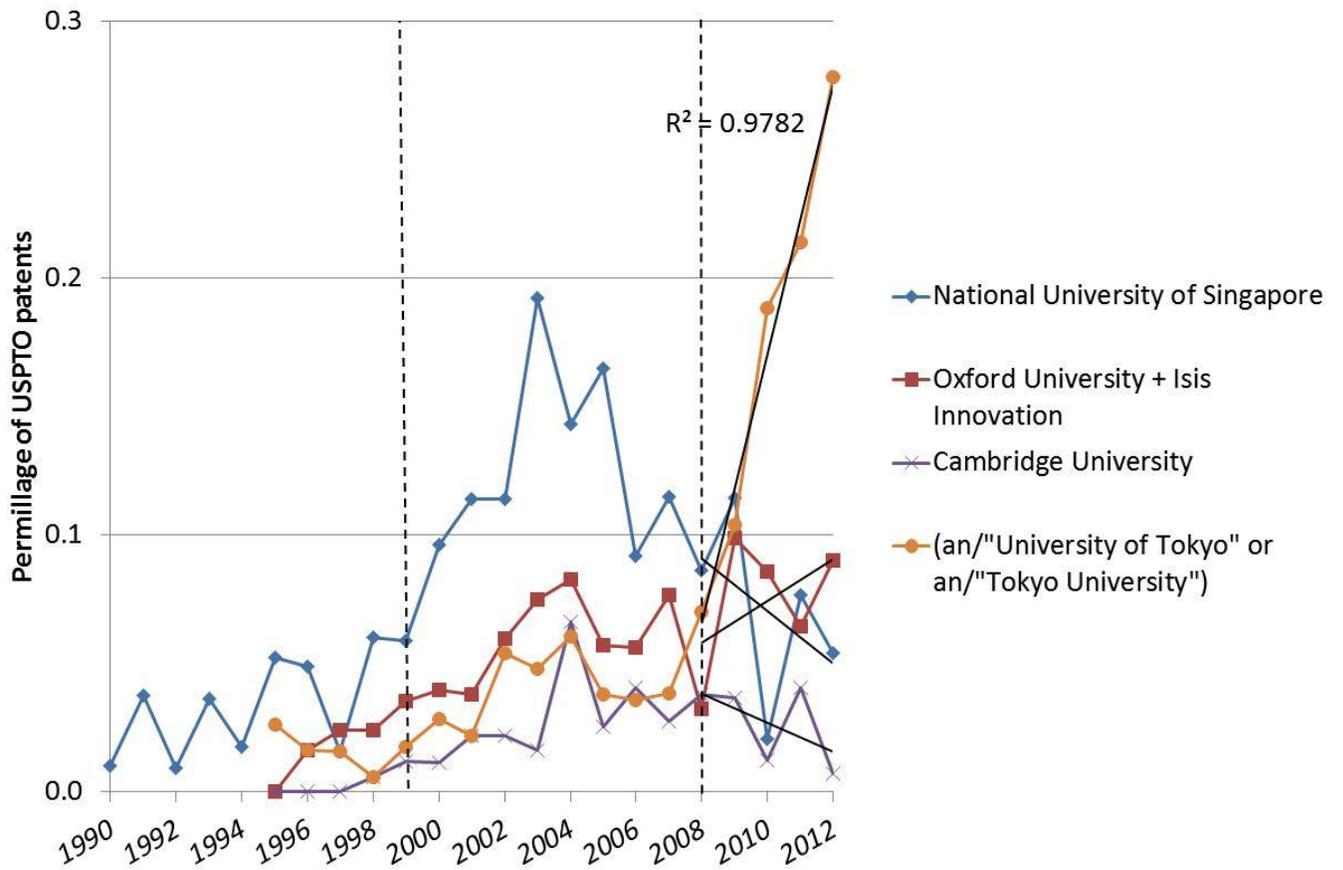

**Figure 3**: Permillage of USPTO patents of four major non-US universities.

Tokyo University shows the most spectacular increase, making its percentage share rapidly equal to that of a major American university. The Japanese government heavily subsidizes and rewards patenting by university staff. At the national level, however, university patenting has no longer increased in Japan since 2006 (Nishimura, 2011; Masashi Shirabe, *personal communication*, Feb. 13, 2013).

A similar effect can be noted for Chinese universities who boosted their university patenting until recently. However, the number of patents held by the National University of Singapore has not increased since an earlier peak in the previous period (Wong & Sing, 2010). Chinese university patenting at USPTO has increased like the pattern at Tokyo University, but three times as much, to a level of 289 patents (1‰) in 2012. In summary, it would require a more detailed analysis—beyond the scope of this rejoinder—to answer the question of which universities precisely carry the recent increases in university patenting at USPTO.

**American patenting in Europe**

The above results raise the question of whether the noted effect is specifically a result of foreign direct investment in a reverse direction, or an effect of globalization in university patenting more generally. For this reason, we checked the database of the European Patent Office (EPO) to see whether similar trends could be found for American universities patenting in Europe. In this database, too, one can witness the effects of the Bayh-Dole Act during the first period and the relative stabilization of university patenting since 2000. However, the increases since 2008 were not significant for the four American universities under study (Figure 4).

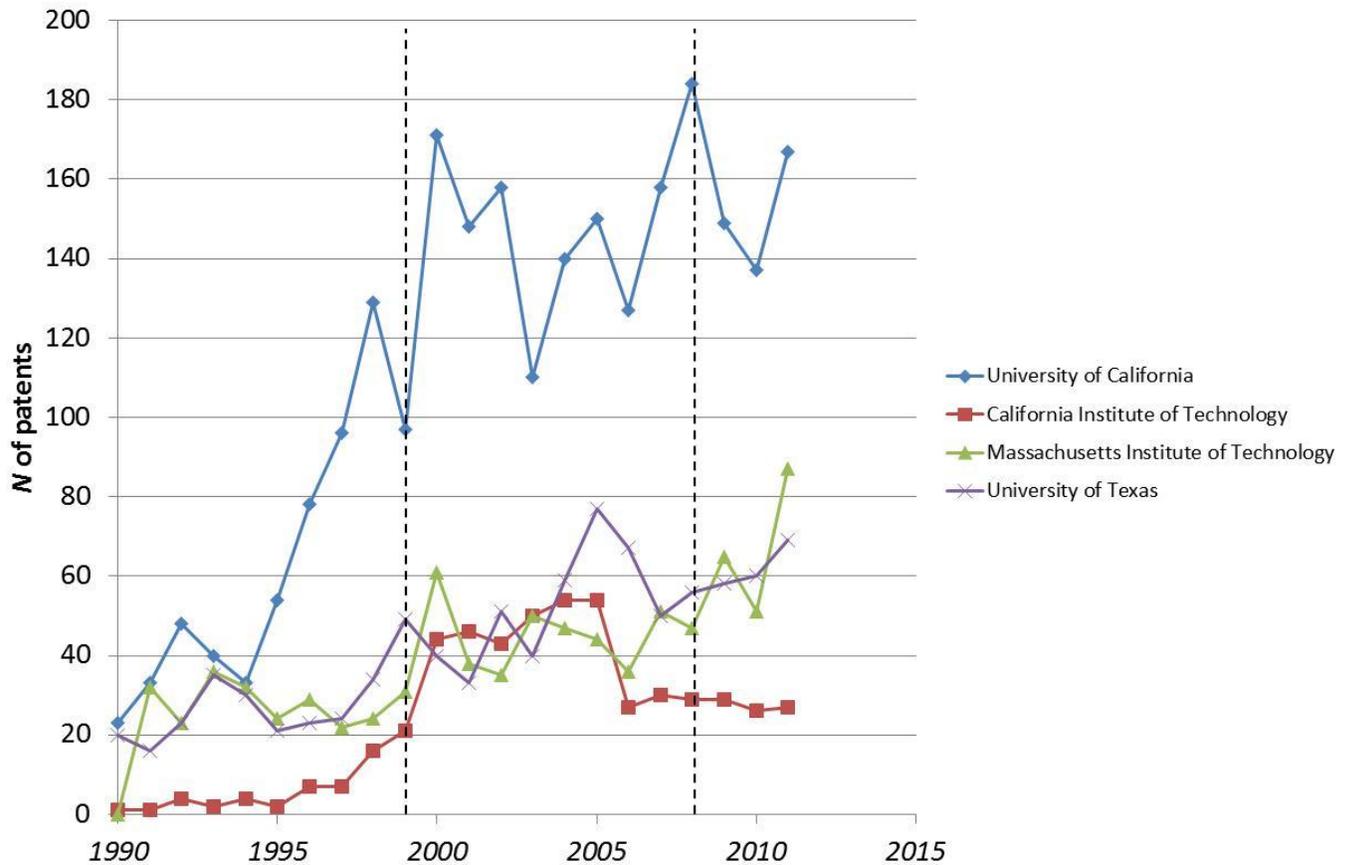

**Figure 4**: Patenting by four major American universities at the European Patent Office EPO.[1]

**Etzkowitz' methodological objections**

In addition to objecting to the interpretation of our scientometric results, Etzkowitz (in press) also raises methodological concerns about using quantitative "numerology" to explain technology transfer as an activity. However, Leydesdorff & Meyer (2010) did not impute the relevance of "the end of the Bayh-Dole effect" for university patenting to a critique of technology transfer. Mowery & Sampat (2004) suggested this connection when they found an upward trend (using data before 2000) which they interpreted as showing the success of the transfer offices. We are aware that Technology Transfer Offices (TTOs) do not control the patenting process.

---

[1] For the years since 2002, one can normalize this data as percentages using aggregate statistics available at http://www.epo.org/about-us/statistics/granted-patents.html.

In his comments, Etzkowitz, in our opinion, advocates using a participant's perspective that focuses on the process more than our outcomes, whereas the scientometric perspective enables us to abstract from specific cases and evaluate developments across a distribution of cases. Of course, one then loses historical detail—the two perspectives stand analytically in orthogonal relationship to each other. However, participants and policy analysts are inclined to focus on "best practices," whereas one can learn more from failures (Leydesdorff, 2003). Our prime concern is motivated by the question about return on investment for universities. Are the policies advocated by Etzkowitz (e.g., 2008) prudent across the board, or very case-specific?

Etzkowitz (in press) argues at length that one should focus not only on patents, but also on startups, etc.[2] Qualitative definitions of participants, however, are often insufficiently precise for scientometric analysis. Since the crisis of 2008, for example, many universities can claim a large number of new startups because starting up a company is an easy option for graduating students to hide their unemployment. In our opinion, one needs statistics that are based on criteria.

The Association of University Transfer Managers (AUTM)—an organization mentioned and acclaimed by Etkzowitz in this context—provides evaluative statistics in its annual reports that have been collected since 2009 in a database (STATT) accessible to members online (at http://www.autm.net/source/STATT/index.cfm?section=STATT; cf. AUTM, 1997; Etzkowitz &

---

[2] Other analysts (e.g., Langford *et al.*, 2006) have also criticized the focus on spin-offs or licensing as too narrow and neglectful of a number of other important paths in knowledge flows.

Stevens, 1998). Let us first note that if the data for patents granted as provided in STATT is used, Figure 1 is virtually replicated since 1978 (see Leydesdorff & Meyer, 2010: 358). This gives confidence in the quality of this data despite its being based on self-reporting in survey research.

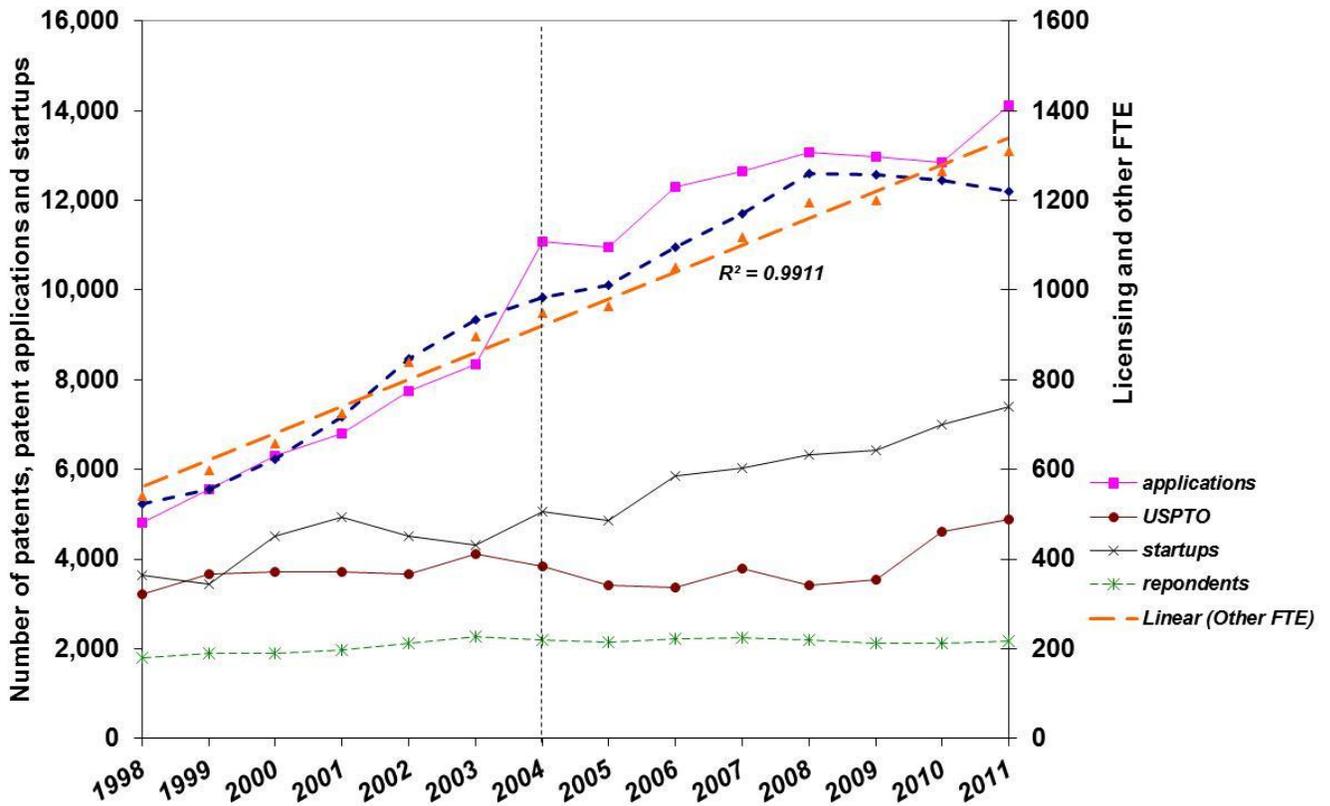

**Figure 5**: Input and output statistics of the Association of University Transfer Managers (AUTM).[3]

Figure 5 shows a number of other parameters based on the tables in STATT, but now plotted longitudinally. First, one can see on the right axis that the numbers of full-time equivalents (FTE), both licensing (in blue) and other tasks (in orange) has increased continuously. The number of patent *applications* (left axis) has increased accordingly. (The curve for the support staff is exceptionally linear with $R^2 > .99$.) The output, however, in terms of *granted* patents (left axis) remained stable from 1998 until the two most recent years, with an average of 3,606.1 (st. dev. = 657.7). In our opinion, this stagnating configuration prevailed until 2004; in more recent years, the number of startups has increased from approximately 500 in 2004 to 739 in 2011.

---

[3] The data for 2012 was not yet available as of the date of our search in STATT on February 2, 2013.

In other words, one could argue that, even on the basis of self-reported data, university patenting has not been very effective in terms of organizations such as Technology Transfer Offices (TTOs): the investments went up, but the output did not follow. Similar effects have been reported elsewhere about incubators and science parks that failed to contribute to the intended outcome (Mustar *et al*., 2006; Clarysse *et al*., 2007). From a sociological perspective (Spiegel-Rösing, 1973; Van den Daele *et al*., 1979), Figure 5 provides an almost perfect illustration of "parametric steering": if more money is spent on a priority, one obtains more activity, but not necessarily the desired outcomes.

We do not wish to deny the many examples that Etzkowitz (in press) provides of successful technology transfer. On the contrary, we agree that universities have multifaceted networks with external stakeholders from industry and government which allow them to contribute to economic development and society at large (Martinelli *et al*., 2008). Our interest is as much motivated by concerns about university-industry-government relations as his, but our approach is more quantitative and information-oriented (Venditti *et al*., in press). However, we do not deny the emergence of new, hybrid organizations at the heart of Triple Helix relations (Meyer *et al*., 2013). While we view these developments as new institutional arrangements, we are also concerned about the validity of a policy analysis that advocates institutional relations and neo-corporatist arrangements as a recipe for regional development *per se* (Etzkowitz, 2008). The Third Mission of universities has obviously evolved since the introduction of the Triple Helix model in the 1990s (Lawton-Smith & Leydesdorff, in preparation; cf. Etzkowitz & Leydesdorff, 1995, 2000).

Leydesdorff & Meyer (2003), among others, argued for developing an evolutionary model of the Triple Helix in which one can distinguish between the institutional arrangements of organizations and the self-organization of interfaces among markets, knowledge development, and normative control. The nonlinear dynamics are structural and potentially resistant to organizational interventions (Leydesdorff, 2006; Leydesdorff & Zawdie, 2010; Kwon *et al.*, 2012). In terms of the three relevant dynamics, patents can be considered as output indicators of the technosciences, input indicators to the economy, and most importantly, as a means for the protection of intellectual property.

**References**

AUTM (1997). *US Licensing Survey 1996*. Deerfield IL: Association of University Transfer Managers.
Clarysse, B., Wright, M., Lockett, A., Mustar, P., & Knockaert, M. (2007). Academic spin-offs, formal technology transfer and capital raising. *Industrial and Corporate Change, 16*(4), 609-640.
Etzkowitz, H. (2008). *The Triple Helix: University-Industry-Government Innovation In Action*. London: Routledge.
Etzkowitz, H. (in press). Mistaking Dawn for Dusk: Quantophrenia and the Cult of Numerology in Technology Transfer Analysis -- A comment to: Leydesdorff L. and Meyer M. (2010). The decline of university patenting and the end of the Bayh-Dole effect. *Scientometrics*.
Etzkowitz, H., & Leydesdorff, L. (1995). The Triple Helix---University-Industry-Government Relations: A Laboratory for Knowledge-Based Economic Development. *EASST Review 14*, 14-19.
Etzkowitz, H., & Leydesdorff, L. (2000). The Dynamics of Innovation: From National Systems and 'Mode 2' to a Triple Helix of University-Industry-Government Relations. *Research Policy, 29*(2), 109-123.

Etzkowitz, H., & Stevens, A. J. (1998). Inching toward industrial policy: The university's role in government initiatives to assist small, innovative companies in the United States. In H. Etzkowitz, A. Webster & P. Healy (Eds.), *Capitalizing knowledge: New intersections of industry and academia* (pp. 215-238). New York, NY: SUNY Press.

Jaffe, A. B., & Trajtenberg, M. (2002). *Patents, Citations, and Innovations: A Window on the Knowledge Economy*. Cambridge, MA/London: MIT Press.

Kwon, K. S., Park, H. W., So, M., & Leydesdorff, L. (2012). Has Globalization Strengthened South Korea's National Research System? National and International Dynamics of the Triple Helix of Scientific Co-authorship Relationships in South Korea. *Scientometrics, 90*(1), 163-175.

Langford, C.H., Hall, J., Josty, P., Matos, S., Jacobson, A. (2006). Indicators and outcomes of Canadian university research: Proxies becoming goals? *Research Policy*, *35*(10), 1586-1598.

Lawton-Smith, H., & Leydesdorff, L. (in preparation). The Triple Helix in the context of global change: Continuing, mutating, and unraveling.

Leydesdorff, L. (2003). A methodological perspective on the evaluation of the promotion of university-industry-government relations. *Small Business Economics, 22*(2), 201-204.

Leydesdorff, L. (2006). *The Knowledge-Based Economy: Modeled, Measured, Simulated*. Boca Raton, FL: Universal Publishers.

Leydesdorff, L., & Meyer, M. (2003). The Triple Helix of University-Industry-Government Relations: Introduction to the Topical Issue. *Scientometrics, 58*(2), 191-203.

Leydesdorff, L., & Meyer, M. (2010). The Decline of University Patenting and the End of the Bayh-Dole Effect. *Scientometrics, 83*(2), 355-362.

Leydesdorff, L., & Zawdie, G. (2010). The Triple Helix Perspective of Innovation Systems. *Technology Analysis & Strategic Management, 22*(7), 789-804.

Martinelli, A., Meyer, M., Von Tunzelmann, G.N. (2008). Becoming an entrepreneurial university? A case study of knowledge exchange relationships and faculty attitudes in a medium-sized, research-oriented university. *Journal of Technology Transfer*, *33*(3) 259–283. 1.

Meyer, M., Grant, K., Kuusisto, J. (2013). The Second Coming of the Triple Helix and the Emergence of Hybrid Innovation Environments. In: R. Capello, A. Olechnicka, G. Gorzelak (Eds.), *Universities, Cities and Regions: Loci for Knowledge and Innovation Creation*. London/New York: Routledge, 193-209.

Mowery, D. C., & Sampat, B. N. (2004). The Bayh-Dole Act of 1980 and University–Industry Technology Transfer: A Model for Other OECD Governments? *The Journal of Technology Transfer, 30*(1), 115-127.

Mustar, P., Renault, M., Colombo, M. G., Piva, E., Fontes, M., Lockett, A., . . . Moray, N. (2006). Conceptualising the heterogeneity of research-based spin-offs: A multi-dimensional taxonomy. *Research Policy, 35*(2), 289-308.

Nishimura, Y. (2011). Recent trends of technology transfers and business-academia collaborations in Japanese universities. *Journal of Industry-Academia-Government Collaboration, 7*(1), 13-16.

Spiegel-Rösing, I. (1973). *Wissenschaftsentwicklung und Wissensschaftssteuerung*. Frankfurt a.M.: Athenaeum Verlag).

Stevens, A. J. (2004). The enactment of Bayh–Dole. *The Journal of Technology Transfer, 29*(1), 93-99.

Van den Daele, W., Krohn, W., & Weingart, P. (Eds.). (1979). *Geplante Forschung: Vergleichende Studien über den Einfluss politischer Programme auf die Wissenschaftsentwicklung*. Frankfurt a.M.: Suhrkamp.

Venditti, M., Reale, E., & Leydesdorff, L. (in press). The Disclosure of University Research for Third Parties: A Non-Market Perspective on an Italian University *Science and Public Policy*, available at http://arxiv.org/abs/1111.5684

Wong, P. K., & Singh, A. (2010). University patenting activities and their link to the quantity and quality of scientific publications. *Scientometrics, 83*(1), 271-294.